# An intuitive introduction to the concept of spatial coherence

H. J. Rabal, N. L. Cap, E. Grumel, M. Trivi

Centro de Investigaciones Ópticas (CONICET La Plata - CIC) and UID Optimo,

Departamento Ciencias Básicas, Facultad de Ingeniería, Universidad Nacional de la Plata,

P.O. Box 3, Gonnet, La Plata 1897, Argentina

**ABSTRACT**

The concept of spatial coherence is usually hard to be understood the first time that it is studied. We propose here a fully intuitive geometric description that does not contain mathematical difficulties and permits to understand how a Young´s Fringes system is obtained with a source not spatially coherent. It is based in a very simple experiment that permits the detection of spatial coherence in a scene. Experimental results are shown.

## 1. INTRODUCTION

The concept of spatial coherence is usually difficult to understand. An introductory approach on the coherence subject was developed by M. L. Calvo [1].

The rigorous study of the spatial coherence has been developed by Van Cittert and Zernike [2]. It states that the normalized degree of coherence is the Fourier transform of the intensity distribution for uncorrelated emitters.

When the interference phenomenon is studied, it usually starts with the calculation of the irradiance in an observation plane due to the superposition of two waves.

In the classical calculation of the field due to plane waves coming from two sources, if they are coherent (they keep a constant phase difference) and monochromatic (have the same frequency), two facts that, when taken rigorously result to be only one as if the waves are strictly

monochromatic their phase difference is effectively constant, then, the waves can be described as [3]:

$$\vec{E}_i(\vec{r},t) = \vec{E}_{0i} sen(\vec{k}_i \cdot \vec{r} - \omega t + \varepsilon_i)$$

$$i = 1,2$$

where $\varepsilon_i$ is the inital phase of the wave.

In any point where both waves overlap, the total field is

$$\vec{E} = \vec{E}_1 + \vec{E}_2$$

And the irradiance is

$$I = \varepsilon.v\left(E_1^2 + E_2^2 + 2\vec{E}_1 \cdot \vec{E}_2\right) = I_1 + I_2 + I_{12}$$

$$I_{12} = 2\varepsilon.v\left(\vec{E}_1 \cdot \vec{E}_2\right)$$

$$I_{12} = 2\varepsilon.v\vec{E}_{01} \cdot \vec{E}_{02}\left(\cos(\vec{k}_1 \cdot \vec{r} - \omega t + \varepsilon_1)\cos(\vec{k}_2 \cdot \vec{r} - \omega t + \varepsilon_2)\right)$$

$$= 2\varepsilon.v\vec{E}_{01} \cdot \vec{E}_{02}\left(\cos(\varphi_1 - \omega t)\cos(\varphi_2 - \omega t)\right)$$

By using the identities

$$\cos(A \pm B) = \cos A \cos B \mp \sin A \sin B$$

$$\langle \cos^2(\omega t) \rangle = \langle \sin^2(\omega t) \rangle = \frac{1}{2}$$

$$\langle \cos(\omega t) \rangle = \langle \sin(\omega t) \rangle = 0$$

$$I_{12} = \varepsilon.v\vec{E}_{01} \cdot \vec{E}_{02} \cos\delta_{12}$$

with

$$\delta_{12} = \left(\cos(\vec{k}_1 \cdot \vec{r} - \omega t + \varepsilon_1) - \cos(\vec{k}_2 \cdot \vec{r} - \omega t + \varepsilon_2)\right)$$

If the field vectors are parallel:

$$I_{12} = 2\sqrt{I_1 I_2} \cos\delta_{12}$$

It is, the total irradiance is found to be [3]:

$$I = 2I_0(1 + \cos \delta_{12}) = 4I_0 \cos^2\left(\frac{\delta_{12}}{2}\right) \tag{1}$$

Where $I_0$ are the irradiance of the two waves and $\delta_{12}$ is the phase difference between them. We assumed that the electrical fields of both waves are parallel.

The phase difference $\delta_{12}$ in the case of incoherent sources is not constant but changes so fast and ordinary detectors cannot detected the interference phenomenon. The irradiance due to two such sources is the sum of the irradiances of each of them.

Nevertheless, when we observe a light source through a pupil composed by two thin slits, we find a fringe system. Why does it happen? How can we justify it?

How is it possible that two (or more) incoherent elements of a source could give rise to nonzero visibility fringes stable in time in spite of the fact that their relative phases are fluctuating randomly?

Each idealized point element of it produces high contrast fringes, but different elements are not supposed to give rise to stable interference patterns.

Then, the only possibility for high visibility to subsist when both point source elements are present should be that the source points are separated by such a distance that makes the individual fringe systems to coincide.

Then, emphasis is exerted in the fact that the existence of measurable fringes visibility is due to the superposition in consonance of multiple fringe systems originating in different source elements that are incoherent between them.

This idea, originally used to calculate visibility in times before the VCZ Theorem was stated , can be exemplified by using two very small and close pinholes very near to the eyes and observing through them outdoor scenarios (see Figure 1). Even if the available light is not monochromatic, fringes can be observed in luminance discontinuities, such as edges, wires or poles, images of the Sun in dew drops or cylindrical surfaces also show fringes with visibility high enough to be discerned. It is easy, then, to figure out that low or zero visibility in extended sources is due to the superposition of shifted fringes systems.

We suggest here a description using elementary trigonometric identities to explain how a Young Fringes system can be obtained from a source constituted by incoherent point sources. The visibility in the fringes with a compound source is found as the coincidence of several shifted fringes systems coming each from every single point. These are added on an intensity basis.

This approach leads in a natural way to the same result for the visibility as the Van Cittert-Zernike Theorem for any arbitrary source distribution.

We use a simple experiment to illustrate this proposal. It is consisting in the observation of a scene through a card with two very small and very close slits.

## 2. SIMPLE EXPERIMENT WITH A NATURAL SCENE

If we observe a point like light source with an optical system limited by two parallel slit apertures, it can be observed that in the image there is Young´s Fringes fringe pattern.

If the source is composed by several incoherent emitters, the observable irradiance are too fast to be detected in the optical range, their average is zero and no fringes are observed.

Nevertheless, if this optical system is pointed to any natural scene, it can be that it appears covered with fringes. As an example, in Figure 1 we show a natural scene (a backyard) though an optical system (the camera) limited by two thin parallel slits. The corresponding Young Fringes that can be observed in the irradiance discontinuities.

How we can solve this contradiction? Why we find fringes?

To look for the reason we are going to consider very simple sources and to ask what happens with the fringes that they give rise.

# 3. TWO MUTUALLY NON COHERENT POINT LIGHTSOURCES

In Figure 2, A and B in plane $\pi$ represent two quasi monochromatic point sources separated by a distance $X_0$. They have the same wavelength $\lambda$ and the same irradiance $I_0$. Narrow slits $P_1$ and $P_2$ are separated a distance $d$. The lens L, with focal distance $f_0$ conjugates the planes $\pi$ and $\pi´$. The distance $z$ is much bigger than $d$ distances and the distance $z´ \sim f_0$.

For each quasi monochromatic source corresponds in plane $\pi$ to the light distribution found in a Young´s Fringes experiment.

The irradiance distributions $I_A$ and $I_B$ in plane due to the sources A and B respectively can be described as:

$$I_A(x) = 4I_0 \cos^2\left[\frac{\pi}{\lambda}\frac{d}{f_0}x\right] \qquad (2)$$

$$I_B(x) = 4I_0 \cos^2\left[\frac{\pi}{\lambda}\frac{d}{f_0}(x - X´_0)\right] \qquad (3)$$

The visibility of the fringes system produced by $I_A$ and $I_B$, when both overlap, is:

$$V = \frac{I_{MAX} - I_{MIN}}{I_{MAX} + I_{MIN}} = \left|\cos\left(\frac{\pi . X´_0 d}{\lambda z'}\right)\right| \qquad (4)$$

The visibility depends on the separation $d$ between $P_1$ and $P_2$ (points that are used in the correlation in the Van Cittert Zernike´s Theorem) as well as the relation between $\frac{X'_0}{Z'}$ and the wavelength.

Imposing a $2\pi$ (or integer multiples of it) shift between both fringes systems, they will be in consonance and the visibility of the composed system will be a maximum. It is due to coincidence *of the fringes* and not to interference between light coming from the different sources.

# 4. CONTINUOUS SOURCE DISTRIBUTIONS

Following the same line of reasoning as before, if there are N discrete point sources with irradiance Ii, located at points xí, the intensity distribution in plane Q-Q´ results:

$$I(x) = \sum_{i=1}^{N} I_i \cos^2[\omega_W(x - x_i)]$$

As is described in [4], for the case of a continuous quasi monochromatic incoherent intensity distribution source, the visibility of the Young fringes systems becomes:

$$I(x) = \int I(x') \cos^2(\omega_x(x'-x)) dx'$$

$$= B + \frac{1}{2} A \cos(2\omega_x x - \delta)$$

with

$$A \cos \delta = \int I(x') \cos(2\omega_x x') dx'$$

and

$$A \sin \delta = \int I(x') \sin(2\omega_x x') dx'$$

$$B = \int I(x') dx'$$

where $I(x´)$ is the density of irradiance per unit length.

Then, the maximum and the minimum irradiances will be, respectively:

$$I_{Max} = B + \frac{1}{2} A; \quad I_{Min} = B - \frac{1}{2} A$$

and the visibility is:

$$V = \frac{I_{Max} - I_{Min}}{I_{Max} + I_{Min}} = \frac{A}{2B}$$

$$V = \frac{\left\{\left|\int I(x')\cos(2\omega_x x')dx'\right|^2 + \left|\int I(x')\sin(2\omega_x x')dx'\right|^2\right\}^{1/2}}{2\int I(x')dx'} \qquad (5)$$

where

$$w_x = \frac{\pi}{\lambda}\frac{d}{f_0} \qquad (6)$$

The visibility is given by the modulus of the normalized Fourier Transform of the intensity density distribution of the source. This is the result of the Van Cittert Zernike´s Theorem. Visibility of the fringes depends on the distance d (coordinates difference) between points P1 and P2 and not in their actual position in front of the lens.

Notice that only elementary calculus and trigonometric identities are used in this description and at every step a clear understanding is easily maintained of their meaning,

To add another experimental verification, a V shaped object (a transparence mask with two bright convergent slits) can be used and observed through the double parallel slits Young´s experiment to observe how the separation of the points of the V produce fringes systems that add their effects or cancel them according to their distance (see Figure 3) giving a periodic (chromatic, as it depends also in wavelength) variation of visibility.

Does visibility fulfill a Babinet Principle? If this approach is correct, then, given a mask M, its complementary mask $\tilde{M}$ should hide those fringe contributions to the image that are themselves complementary to those given by M.

If it, as shown in equation 5, is the modulus of the normalized Fourier Transform of the source distribution, it could be expected that complementary sources would give rise to similar fringes, but contrast reversed.

Figure 3 shows the result of the experiment when the source is a uniform distribution covered with a thin slit. Young´s Fringes can be seen against the uniform background. Intuitively, it could be thought as the dark slit covering a fringes system that would cancel the fringes generated by a set of points in the background. If this cancellation is prevented by the slit, the fringes can be observed and are contrast reversed with respect to those of the complementary screen.

If P is the transmittance in intensity function of a certain binary screen, then $\bar{P}=1-P$ describes the transmittance of its complementary screen. The visibility of the Young´s Fringes obtained with the latter, as a function of the distance *d* between slits, consists in a delta distribution in the origin of frequencies minus the Fourier Transform of P. The change of sign indicates contrast reversal. So, Babinet´s Principle holds but with a minor change in its interpretation.

Figure 4 shows the results obtained using as object a curved slit and its complementary.

## 5. CONCLUSIONS

When a single point-like source is observed though a two narrow slits, we obtain a Young fringes pattern, where the contributions of each aperture are added on a field basis.

For the spatial extended light sources, high visibility Young fringes can still be observed if every point of it gives rise to a fringes system that coincides with the produced by the others. In this case the addition of these elementary contributions is in intensity.

For the systems to coincide and obtaining a good visibility result there should not exist source points very near that spoil the others visibility. This only happens when the source exhibits spatial discontinuities (i.e. When the Fourier Transform is not a Dirac´s Delta distribution). It is the presence of source discontinuities that gives rise to the visibility predicted by the Van Cittert-Zernike Theorem.

This is also true when the source is a uniform field and there are isolated discontinuities (dilute dark object on bright uniform field). So that visibility behaves as fulfilling a Babinet like complementarity property, although giving fringes that are contrast reversed between a mask and its complementary.

A Holodiagram description of some of these phenomena can be found in [4], where this approach is extended to somewhat more complex sources distributions.

When the source distribution is not entirely contained in a plane perpendicular to the optic axis, the calculation turns out to be a little more involved but still can be described using this geometric approach. [4].

Small departures in experimental visibility observation can be expected if the slits are no extremely thin.

## ACKNOWLEDGEMENTS

This research was supported by grants of University of La Plata, ANPCyT, CONICET and CICPBA (Argentina)

**FIGURES CAPTIONS**

**Figure 1:** A natural scene showing Young´s fringes (slits horizontal)

**Figure 2:** Young fringes pattern obtained from two point sources mutually non coherent.

**Figure 3:** Fringes produced by a V shaped double slit used as incoherent object and using two parallel slits in front of the objective, as in Young´s experiment. Two different V were included and a vertical single slit in the middle for comparison. Notice the periodic variation and contrast reversals in visibility in the vertical direction in the Vs.

**Figure 4:** The fringes obtained with an object in the shape of a curved slit and its complementary one showing Babinet´s Principle.

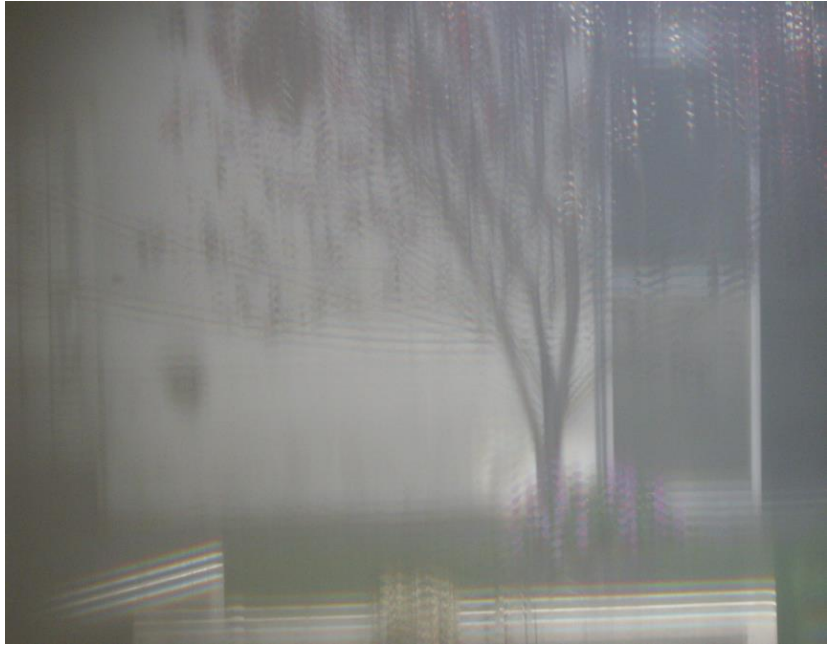

Figure 1 A natural scene showing Young´s fringes (slits horizontal)

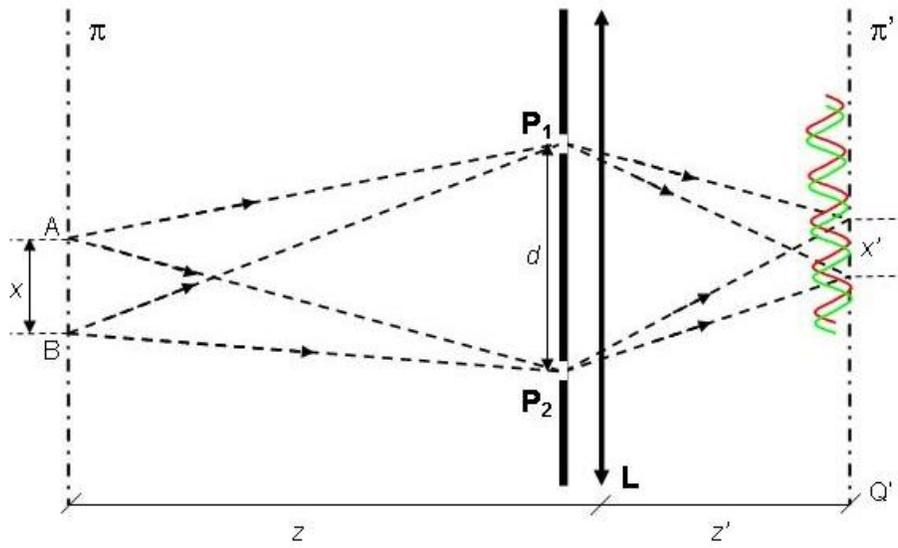

**Figure 2.** Young fringes pattern obtained from two point sources mutually non coherent.

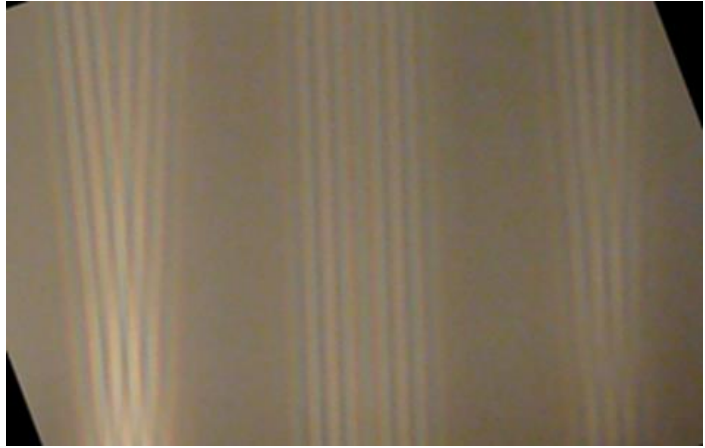

Figure 3: Fringes produced by a V shaped double slit used as incoherent object and using two parallel slits in front of the objective, as in Young´s experiment. Two different V were included and a vertical single slit in the middle for comparison. Notice the periodic variation and contrast reversals in visibility in the vertical direction in the Vs.

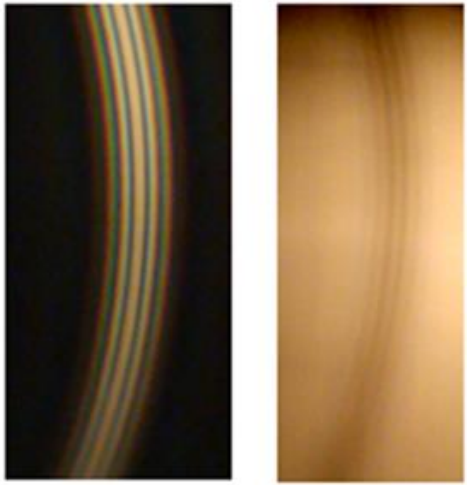

*Figure 4: The fringes obtained with an object in the shape of a curved slit and its complementary one showing Babinet´s Principle.*